\def\cm2{cm$^{-2}$}

\def\c2{C~{\sc ii}}
\def\c4{C~{\sc iv}}
\def\fe2{Fe~{\sc ii}}
\def\fe3{Fe~{\sc iii}}
\def\mg1{Mg~{\sc i}}
\def\mg2{Mg~{\sc ii}}
\def\si2{Si~{\sc ii}}
\def\si4{Si~{\sc iv}}
\def\al2{Al~{\sc ii}}
\def\al3{Al~{\sc iii}}
\def\o1{O~{\sc i}}
\def\n1{N~{\sc i}}
\def\h1{H~{\sc i}}

\def\approxlt{\mathrel{\spose{\lower 3pt\hbox{$\sim$}}
        \raise 2.0pt\hbox{$<$}}}
\def\approxgt{\mathrel{\spose{\lower 3pt\hbox{$\sim$}}
        \raise 2.0pt\hbox{$>$}}}

\documentclass[article]{has40}
\usepackage{graphicx}
\usepackage{amssymb}
\tabletypesize{\normalsize}  

\def\plotone#1{\centering \leavevmode
\includegraphics[width=.95\columnwidth]{#1}}

\def\plotone#1{\centering \leavevmode
\includegraphics[width=.95\columnwidth]{#1}}

\shortauthors{Osborn}
\shorttitle{Man versus Machine}

\begin{document}
\large    
\pagenumbering{arabic}
\setcounter{page}{63}

\title{MAN VERSUS MACHINE: 
\vspace{1pt}eye estimates in the age of digital imaging}

%
%
\author{{\noindent Wayne Osborn{$^{\rm 1,2}$} }\\
\\
{\it (1) Central Michigan University, Mt. Pleasant, MI, USA\\
(2) Yerkes Observatory, Williams Bay, WI, USA} 
}

%
%
\email{Wayne.Osborn@cmich.edu }


\begin{abstract}
Astronomical observing has been greatly simplified by the development and implementation of digital imaging techniques and remote observing.  Aperture photometry of CCD data permits photometric measurements to be made routinely with uncertainties of a few hundredths of a magnitude or better. The question of whether there is still a place in modern observational astronomy for simple eye estimates of brightness is considered.  Examples of recent uses of eye estimates are presented. Suggestions for when eye estimates should be avoided and when they are still worthwhile are offered.  The reactions to these suggestions by the conference audience are summarized.
\end{abstract}

\section{Introduction}
Advances in technology have radically changed the manner by which optical observational astronomy is carried out.  Long nights at the telescope making visual or photographic observations have generally been replaced.  Now observers obtain digital images using CCD detectors, with the telescope usually operated remotely and often located at a distant site with good weather and seeing  or even in space.

Photometry in particular has greatly benefited from CCD observations.  In contrast to the eye or photographic emulsions, the CCD responds linearly with brightness.  The digital nature of the images allows them to be manipulated to improve the observations  such as by removing defects, stacking multiple frames to increase signal to noise, or using image subtraction techniques for photometry in crowded fields.  In most cases routine aperture photometry easily produces magnitudes with errors of only a few percent.   One can therefore ask if there still a place in modern observational astronomy for measuring an object's brightness  through simple eye estimates relative to one or more stars of known magnitude.

\section{Is the eye estimate technique still of value?}
It is a fact that eye estimates of brightness are still being done.  One example is seen in the observations submitted by the members of the American Association of Variable Star Observers.    Figure 1 presents recent AAVSO data for S Her showing that many of the observations are visual ones. Presently about 20$\%$ of the observations submitted to the AAVSO are visual estimates (M. Templeton, private communication), but this statistic is somewhat misleading in that short-period variables generate large numbers of CCD observations while visual observations are still a major component of the data for the longer period variables.  

A second place where eye estimates are still in use is following the brightness changes of objects on archival photographic plates.  Two recent examples can be found in Semkov \textit{et al.}  (2012) and in Osborn, Kopacki and Haberstroh (2012).  Appropriate for this conference, Figure 2 shows the phased light curve of a globular cluster RR Lyrae variable determined from eye estimates of plates obtained at the Michigan State University Observatory 1970-1980.

\begin{figure*}
\centering
\plotone{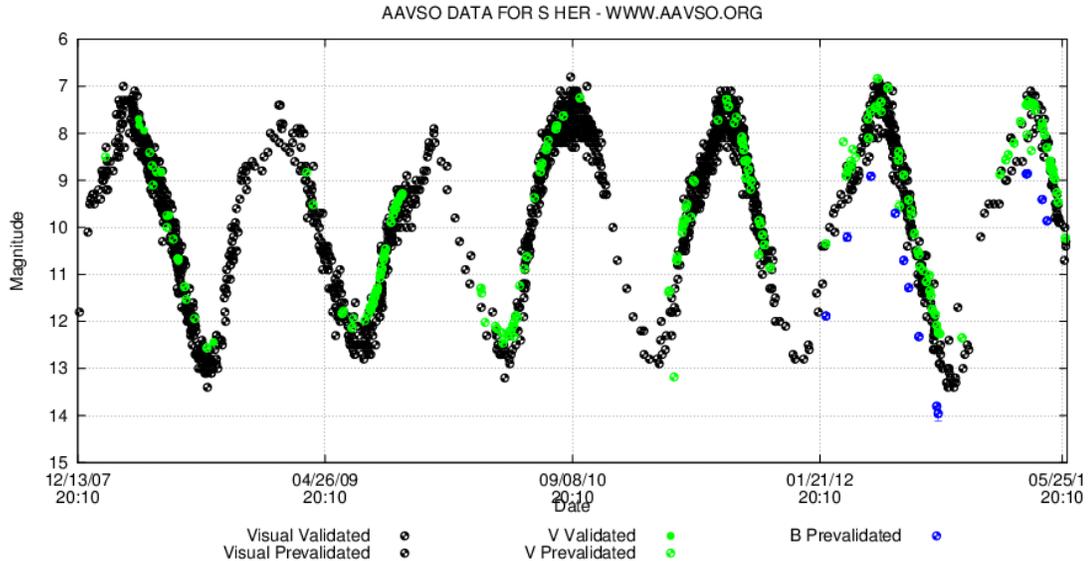}
\vskip0pt
\caption{Fig. 1.  AASVO visual-band data for S Her.  Magnitudes from CCD frames are shown in green, visual estimates in black.  The visual data are important in defining the light curve.}
\label{o1039}
\end{figure*}

Thus, eye estimates \underline{are} presently in use for magnitude determinations, but the issue is if this technique is still worthwhile.  Can data derived from eye estimates yield good science?  Shouldn't all eye estimates be replaced by digitally-based observations?  Let's consider these questions by looking at present astronomical uses of eye estimates. 
		
\section{Present uses of eye estimates and their value}
The eye-estimate technique is in common use in several areas of modern astronomy.  These areas will be briefly identified followed by my opinions on whether or not the data so derived is scientfically useful.  I will invite comments from others in the audience on these opinions at the conclusion of this presentation.  The use of eye estimates in "discovery-type" observations, such as finding that a dwarf nova is in outburst, will be ignored although these are often of great value. 

\subsection{Applications where eye estimates \underline{should be avoided}}
Light curves from visual estimates for eclipsing binaries, RR Lyraes, Cepheids and other objects with relatively short periods are not worth the effort.  Such exercises may be of use as a pedagogical exercise (Turner 1999), but the light curves derived are of too poor a quality to be useful scientifically.  Such observations not only have large errors, light curve modeling requires observations in well-determined (usually standard)  passbands.

More common than light curve determinations for short-period variables are observations of times of minima for eclipsing binaries and times of maxima for RR Lyrae and Cepheid variables.  Such timings are useful for period change studies, but data derived visually, and even photographically, typically have uncertainties many times what can be achieved from photoelectric or CCD observations which can readily be obtained.  Furthermore, there is also the tendency for visual observers to systematically estimate such times too late or to be influenced by the time predicted by the ephemeris.  A good example is given by Zhu \textit{et al.} (2012); the O-C diagram from his paper is reproduced with permission in Figure 3. 

\begin{figure*}
\centering
\plotone{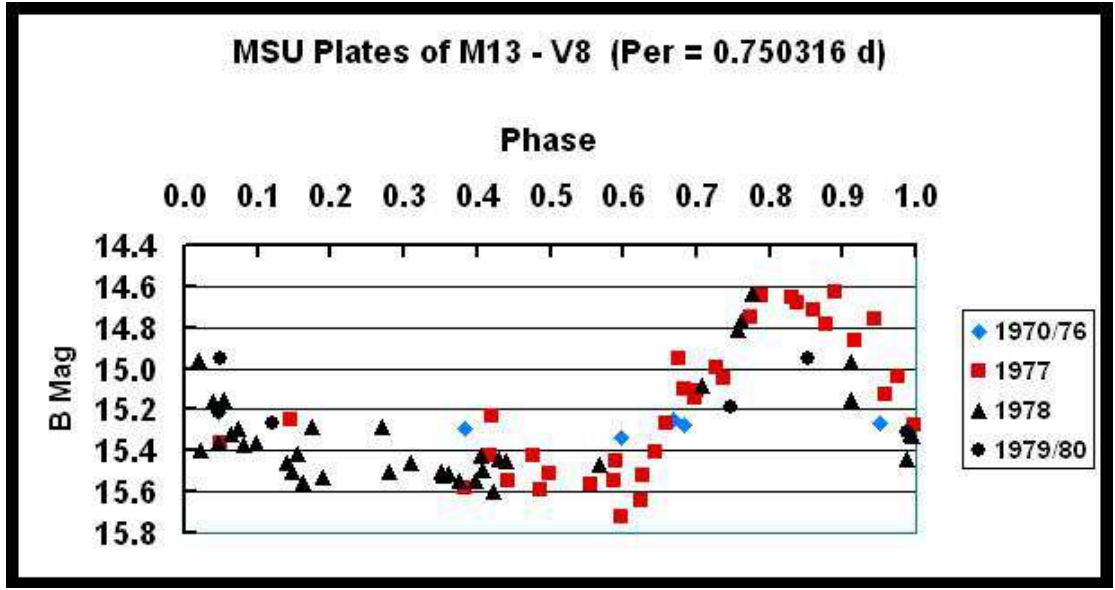}
\vskip0pt
\caption{The phased light curve of M13 Variable 8 from eye estimates of MSU plates.}
\label{o1039}
\end{figure*}

\begin{figure*}
\centering
\plotone{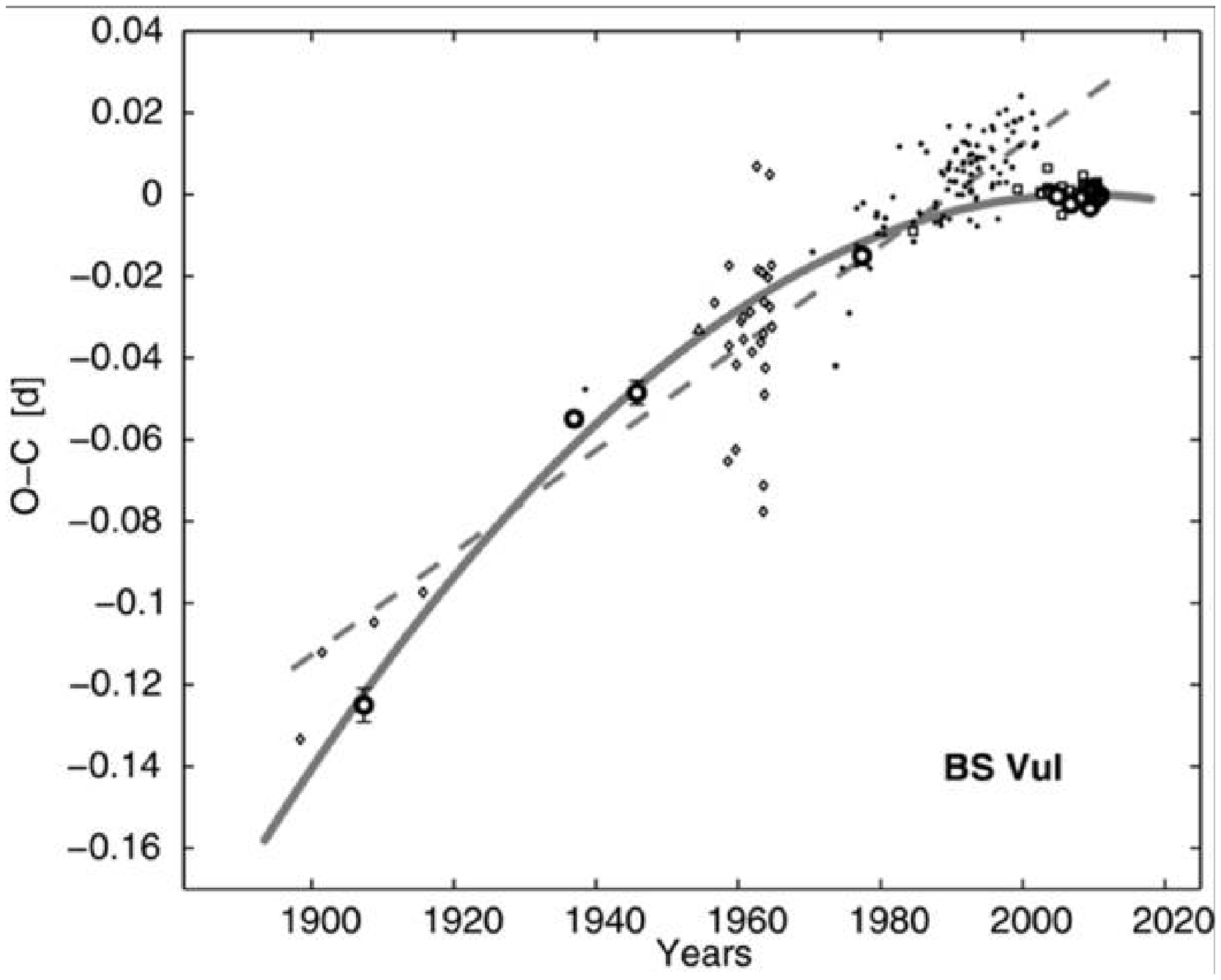}
\vskip0pt
\caption{An O-C diagram for BS Vul from Zhu \textit{et al.} (2012).  The small points after 1980 are values based on visual data.  They show large scatter as well as systematic deviations from the CCD- and photoelectric-based data for this time period.}
\label{o1039}
\end{figure*}

\subsection{Applications where eye estimates are \underline{useful in certain cases}}
Currently, variables with good amplitudes and fairly long periods are routinely monitored by visual estimates as shown by the significant amount of such data for these stars submitted to the AAVSO.  Much of this monitoring may soon be replaced by automated surveys.  Nevertheless, I propose that data from such eye estimates will continue to be of value in certain situations.  

One case is the continuing visual observation of the so called AAVSO legacy variables.  These are stars of special interest because they have been well observed over periods of up to a century or more.  Continued monitoring of these objects in the same manner, i.e. visually, will generate useful long-term self-consistent datasets.  A second case involves the very brightest variables.  In general, the variations of such objects are most easily followed by eye as they are too bright for routine CCD monitoring.  Examples of these two cases, again using AAVSO data, are presented in Figure 4 and Figure 5.

\begin{figure*}
\centering
\plotone{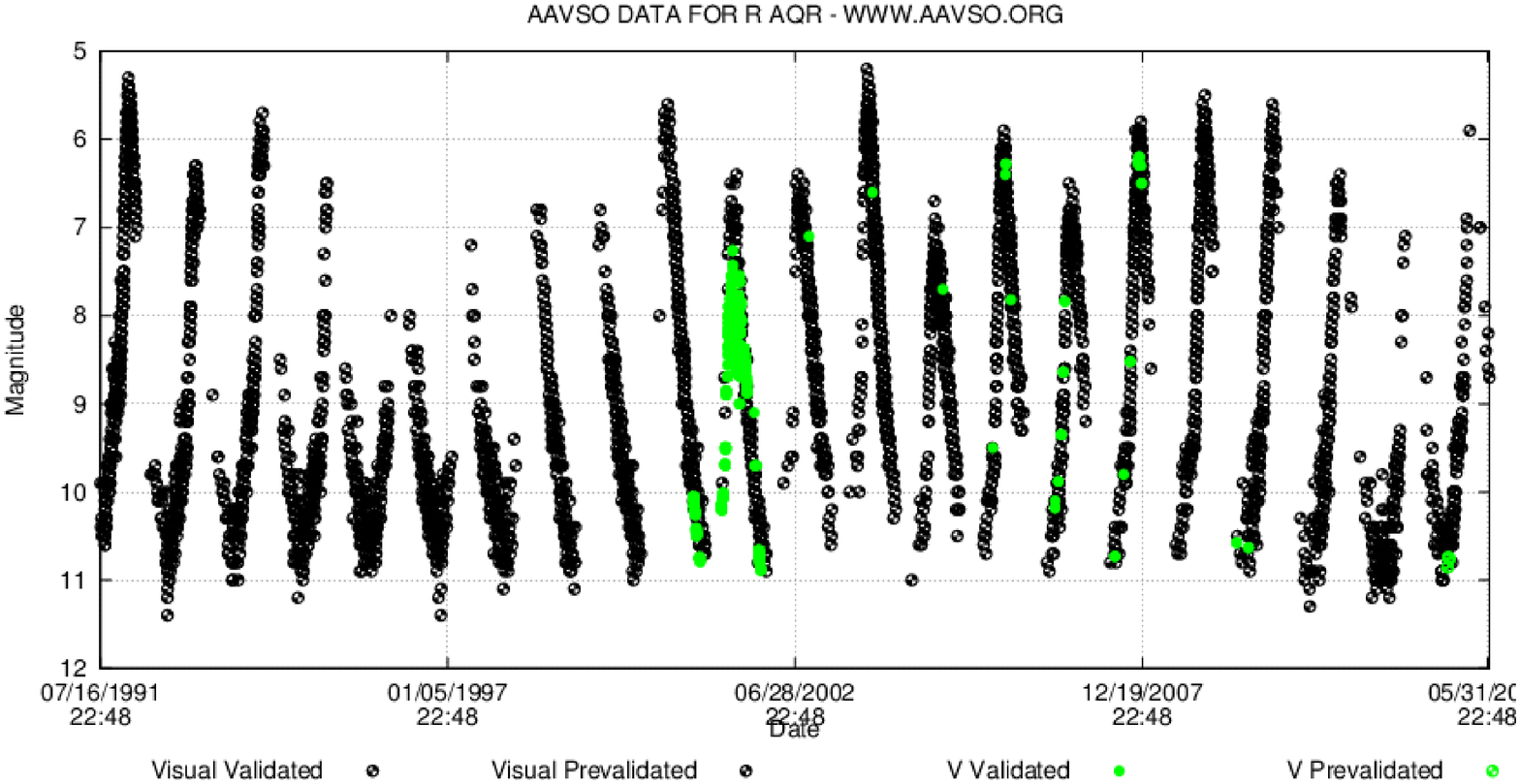}
\vskip0pt
\caption{AASVO data for the legacy long-period variable R Aqr.  The value of the visual observations (in black) for studying the long-term behaviour is obvious.}
\label{o1039}
\end{figure*}

\begin{figure*}
\centering
\plotone{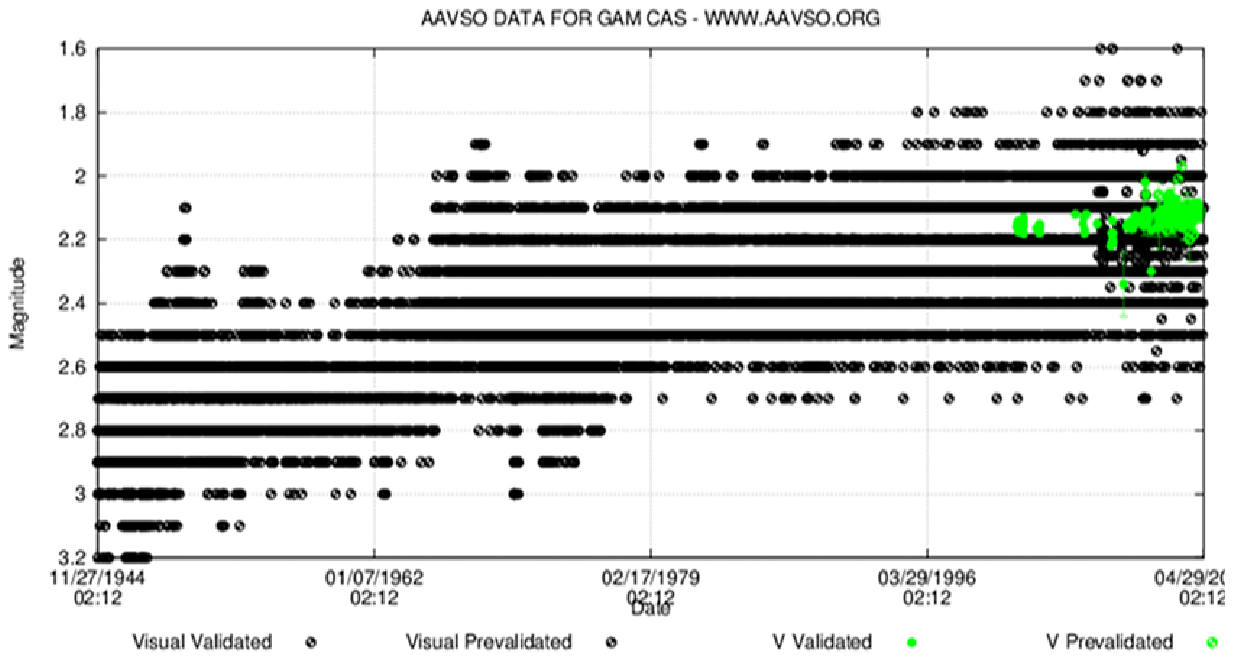}
\vskip0pt
\caption{AASVO data for the bright variable $\gamma$ Cas (V = 2.2).  Despite the large scatter, the visual data (in black) show a slow brightening trend.}
\label{o1039}
\end{figure*}

Another application of eye estimates of scientific value involves the derivation of photometric data from archival photographic plates.  CCD-based data are much more accurate, but the archival photograpic material covers epochs that can't be reproduced.  While it is true that plates can be digitized and then treated similarly to CCD frames, in practice this is time consuming and the reductions can be complicated.  Digitization may not be worth the effort.  This is especially true in cases of crowded fields and when dealing with poorly exposed plates.  In these cases eye estimates are capable of fairly quickly producing data of sufficient accuracy for scientific use.  An example is provided by studies of the variable stars in globular clusters.  Figure 6 shows the cluster M13 (from an MSU plate) with one of the RR Lyrae stars, Variable 8, indicated.   Figure 7 and Figure 8 show phased light curves for Variable 8, the first from photometry of CCD images obtained in 2000 and the other from eye estimates of MSU plates taken in the 1970's.  Fitting such photographic data to an accurate CCD light curve yields epochs of maxima of sufficient accuracy for period change studies (see, for example, Rabidoux \textit{et al.} 2010 and Chicherov 1997).  Eye estimates using photographic material are also commonly used to investigate pre-discovery or long-term photometric behaviour of novae (examples are the studies of Schaefer 2010 and of Jurdana-Sepic and Munari 2008).

\begin{figure*}
\centering
\plotone{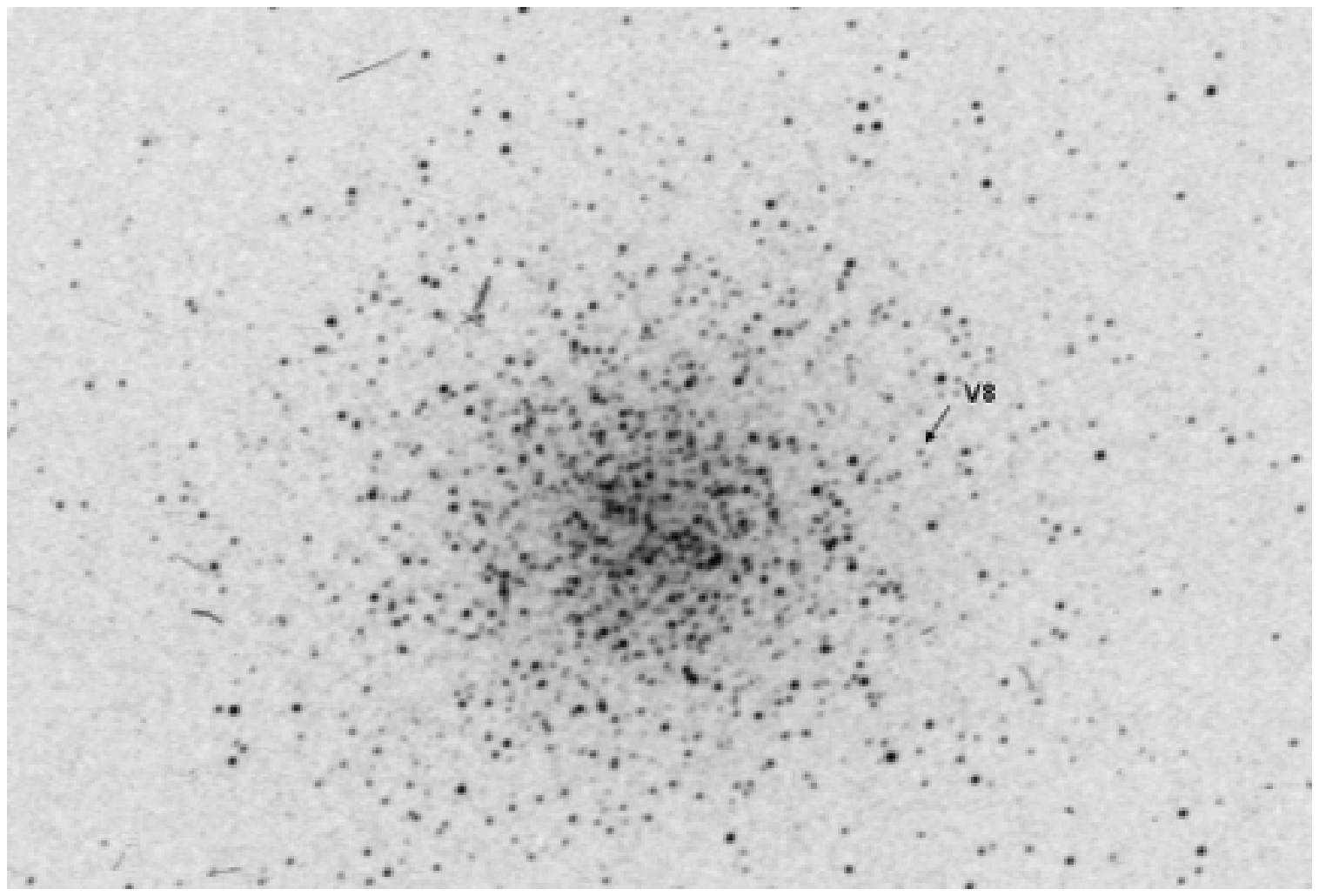}
\vskip0pt
\caption{The globular cluster M13 from a scan of an MSU plate. Variable 8 is indicated.}
\label{o1039}
\end{figure*}

\begin{figure*}
\centering
\plotone{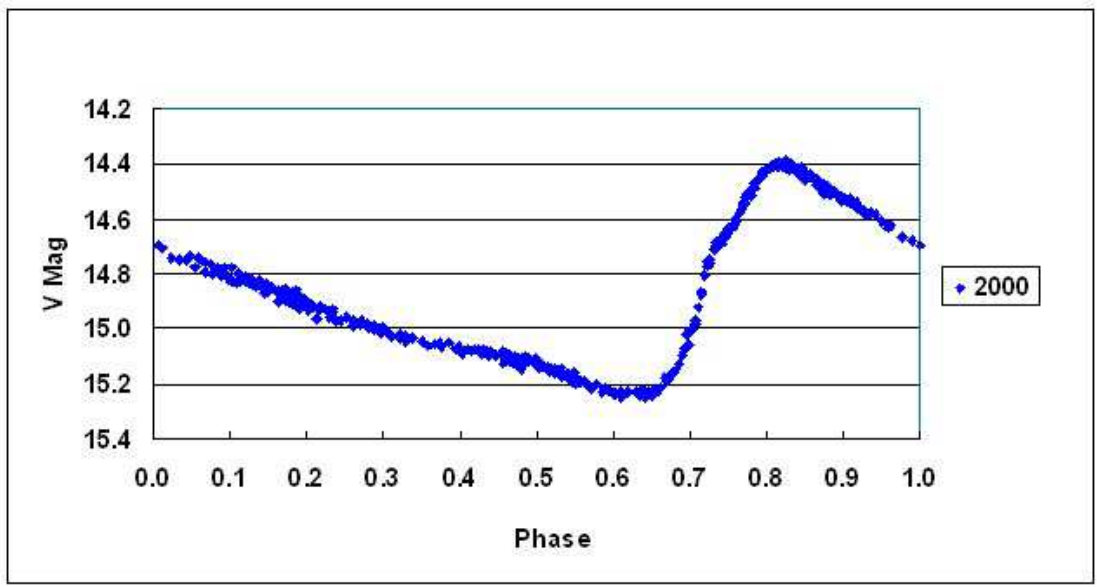}
\vskip0pt
\caption{V light curve for the RR Lyrae variable M13-V8 from CCD images.}
\label{o1039}
\end{figure*}

\begin{figure*}
\centering
\plotone{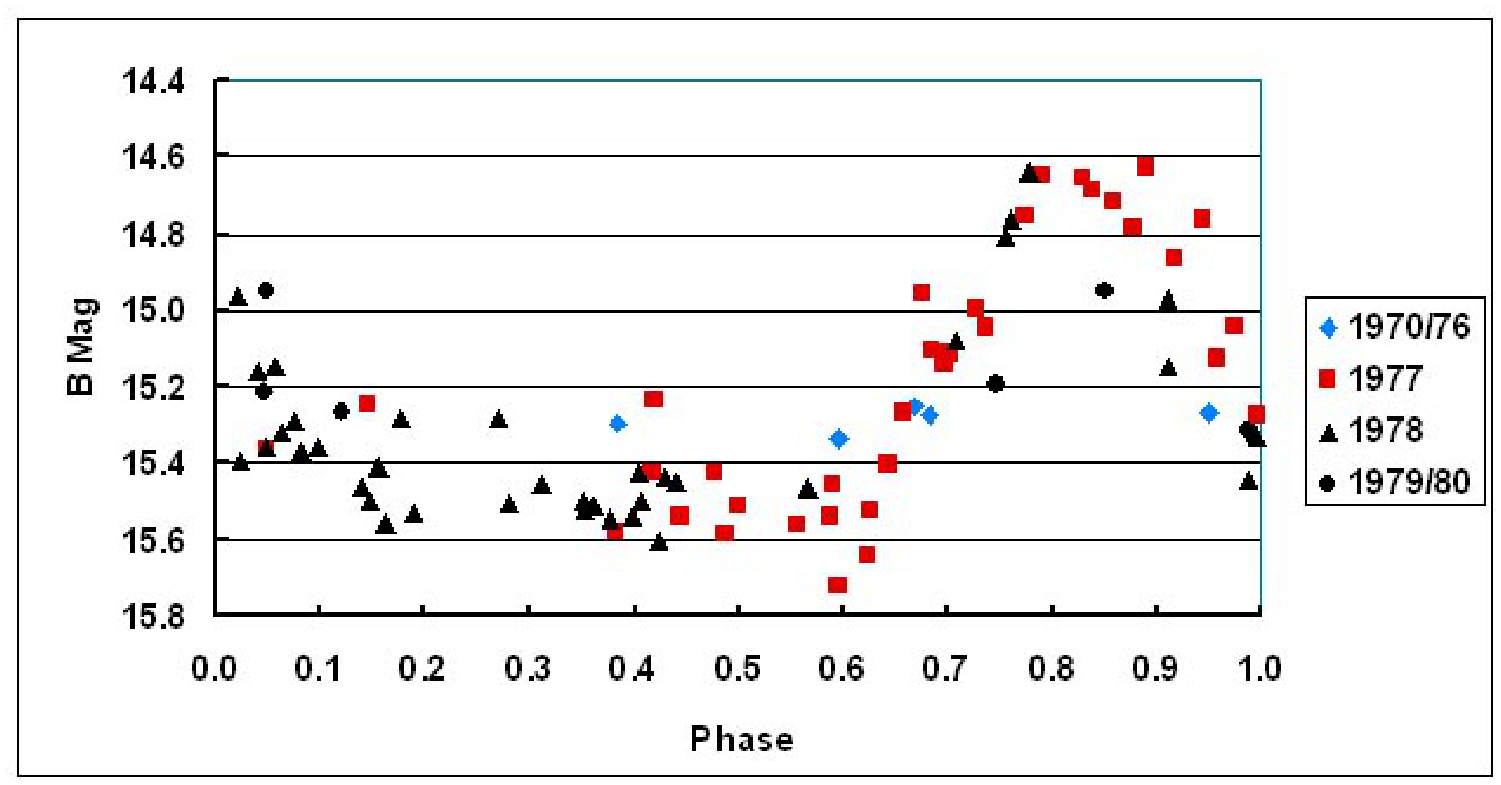}
\vskip0pt
\caption{B light curve for the RR Lyrae Variable M13-V8 from eye estimates of  photographic plates.}
\label{o1039}
\end{figure*}

\newpage
\subsection{An application where eye estimates probably \underline{should be used}}	
One area where eye estimates are probably preferred is for determinations of integrated magnitudes and brightness changes for naked-eye comets.  Such observations are generally done visually and not digitally for several reasons:  comets are extended sources, the observations often must be made at high airmass (and thus below the altitude limits for computer-operated telescopes) and observing conditions can involve bright or poor skies.  This application is illustrated in Table 1 which shows some total-magnitude and coma-diameter observations of Comet C/2012 F6 (Lemmon).  Only one of the measures was done with a CCD.  It is not clear that digital observations could produce better data without a significant amount of effort.

\begin{flushleft}
\begin{deluxetable*}{llcll}
\tabletypesize{\normalsize}
\tablecaption{Total-magnitude and coma-diameter estimates for Comet C/2012 F6 (Lemmon){\tablenotemark{a}}}
\tablewidth{0pt}
\tablehead{ \\ \colhead{2013 Date}   & \colhead{Mag} &
       \colhead{Coma} & \colhead{Observer} &
  \colhead{Notes}   \\
}

\startdata

May 11.13& 6.4& 8' &J. J. Gonzalez, Spain& 10x50 binoculars\\
May  8.78& 6.9& 6' &K. Yoshimoto, Japan&20x100 binoculars\\
May  4.15& 6.2& 6' &J. J. Gonzalez, Spain& 25x100 binoculars \\
May  2.81& 6.8& 3'.8& K. Yoshimoto, Japan& 0.16-m reflector+V filter CCD\\
Apr. 29.35& 6.1&5' &A. Amorim, Brazil&10x50 binoculars, altitude 16 deg; moonlight\\
Apr. 1.35 & 6.0:& 3' &A. Amorim, Brazil & 0.07-m refractor; altitude 5 deg\\ 
Mar. 14.41& 4.3& -- &D. Seargent, Australia & naked eye\\
Mar. 7.42 & 4.4& -- &D. Seargent, Australia & naked eye\\
Mar. 2.95& 5.2& 8'  &A. Amorim, Brazil& 10x50 binoculars\\
Feb. 24.96& 5.2& 8' &A. Amorim, Brazil& 10x50 binoculars; light pollution, moonlight

\enddata
\tablenotetext{a}{Table developed from data published by the \textit{International Comet Quarterly} (see \\
http://www.icq.eps.harvard.edu/CometMags.html) }
\end{deluxetable*}
\end{flushleft}

\newpage
\section{Conclusions}
I believe I have demonstrated that there is still a place in modern observational astronomy for eye estimates of brightness.  Applications where this technique can produce data of scientific value include observations of comet magnitudes, measures on archival photographic plates and the monitoring of variable stars in special cases.  There may be other valid uses that have not been examined, such as in measuring light pollution.   It is suggested that eye estimates should be avoided in favor of filtered CCD or photoelectric observations for determining light curves of eclipsing binaries and short period pulsating variables, such as RR Lyraes and Cepheids, as well as for timing minima and maxima of such objects. 

The suggestions in Section 3 are based significantly on my personal experience.  Members of the audience are encouraged to offer their own views on the use and value of eye estimates.

\subsection{Comments and reactions by the members of the audience}

\noindent \textit{D. Welch}:  First, let me emphasize that the real-time discovery aspect of certain types of visual observations is still a very important role. While the possibilityof computer-based expert systems performing the same function is real, it is likely that visual alerts will maintain the pre-eminence in timeliness, low false-positive rates, and
low cost for some time to come.

On the other hand, there are some assumptions in the earlier discussion with with I disagree.  For instance, we have known what fraction of legacy long-period variables have irregular variations in their O-C diagrams (almost all) and which have "secular" variations which might be associated with expected evolutionary changes. The unstated assumption is that if we 
keep following these objects, we will learn something of additional value. But an equally likely -- or even more probable -- outcome is that we've seen what there is to see and we will just be refining the same statistic. Of course, if we stop observing them, it is certain that we won't learn more! But they will continue to be observed by surveys, and so claiming that
visual observations of legacy LPVs has significant discovery potential is misleading, in my opinion. Also, the competitiveness 
of the visual technique for bright variables is fading with the operation of the  BRITE Constellation satellites (Kushnig \textit{et al.} 2009) and the AAVSO's Bright Star Monitor program (see http://www.aavso.org/bright-star-monitor-stations).

I would also argue that the value of photographic archives is decreasing with time.  
There was no CCD imaging prior to about 1985 and few photographic plates prior to about 1890. 
As the decades pass, larger and larger percentages of the time coverage will be available 
from CCD imaging, and at higher precision.  It won't be long before the span of CCD imaging 
will be long enough to investigate the period changes that have traditionally required 
access to photographic plate archives.  Regarding photometry in globular clusters, the 
total amount of effort required to analyse globular cluster CCD images is VERY low. I would 
argue that it requires more time to do the visual estimates of the variables on the plates!
Still, if the plates exist and people are motviated to measure them, that is a net gain.

Finally, visual estimates of comet magnitudes may be all that is available, but that doesn't 
mean that they are correct. Indeed, one would only be sure of that if one had digital images 
to compare against.  And I have to point out that visual measurements of light pollution from 
"faintest visible stars", which you mentioned in passing, are not estimates of light pollution 
itself but rather a combination of light pollution plus transparency. Sky Quality 
Meters\footnote{Full disclosure by D. Welch: Welch is a co-inventor of the Sky Quality Meter.}, all-sky
cameras, and the Dark Sky Meter phone apps are all modern ways to digitally measure light pollution 
(see, for example, Birriel, Wheatley \& McMichael 2010). \\

\noindent \textit{H. Smith}:  D. Welch rightly noted that -- provided CCD surveys continue and expand -- the importance of existing 
photographic and visual archives will decrease over time.  However, I would point out that the decay time for the utility of such
archives is not short on a human timescale.  Several decades will need to pass before CCD surveys can approach the
time coverage of the photographic data that now extend more than a century into the past.  The value of such
extensive coverage depends, of course, on the timescale of the phenomena being observed. As examples, for RR Lyrae 
variables it is not yet clear whether or not there are significant period change events or other unusual modulations that require time intervals of more more than a few decades to become apparent in individual stars.  There are, however, suggestions that
such phenomena do occur.  Observed period changes for many RR Lyraes stars show $O-C$ variations occur that are large enough to be easily within the range measurable when photographic and visual data are included. Such archival observations therefore remain important for understanding such changes. Eventually CCD time-coverage should catch up, but we are not there yet and I wouldn't 
want to wait 50 years so as to be able to rely solely upon CCD data to address such questions.

That being said, the day is probably soon coming when the role of routine photometric monitoring of variable stars will be taken over by CCD surveys, though, as Doug stated, discovery alerts based on visual monitoring may remain important.  What some have called the romantic era of visual observing by AAVSO members is probably drawing to a close for most purposes. As an AAVSO member for many years, I regard that with some regret, but I certainly like the new results that are emerging from CCD surveys. 

\section{Acknowledgements}
Thanks are due to the AAVSO and its many observers worldwide for making possible Figures 1, 4 and 5 of this paper.  I also thank Dr. L.-Y.  Zhu for permission to reproduce Figure 3.


\begin{thebibliography}



\bibitem[Birriel et al.(2010)]{birriel}~Birriel, J.; Wheatley, J. \& McMichael, C.. 2010, JAAVSO, 38, 132

\bibitem[Chicherov(1997)]{chich}~Chicherov, A. V.  1997, Astron. Letters, 23, 600

\bibitem[Jurdana-\u{S}epi\'{c} \& Munari(2008)]{jurd}~Jurdana-\u{S}epi\'{c}, R. \& Munari, U. 2008, IBVS, no. 5839, 1

\bibitem[Kuschnig et al.(2009)]{kusch}~Kuschnig, R.; Weiss, W. W.; Moffat, A.; \& Kudelka, O.  2009,  ASP Conference Series, Vol. 416, {Solar-Stellar Dynamos as Revealed by Helio- and Asteroseismology: GONG 2008/SOHO 21}.  Ed.  M. Dikpati, T. Arentoft, I. Gonz\'{a}lez Hern\'{a}ndez, C. Lindsey \& F. Hill (San Francisco: Astronomical Society of the Pacific), p. 587

\bibitem[Osborn et al.(2012)]{osborn}~Osborn, W.; Kopacki, G. \& Haberstroh, J. 2012, Acta Astr., 62, 377

\bibitem[Rabidoux et al.(2010)]{rab}~Rabidoux, K.; Smith, Horace A.; Pritzl, B. J.; \textit{et al.} 2010, AJ, 139, 2300

\bibitem[Schaefer(2010)]{schaefer}~Schaefer, B.  2010, ApJS, 187, 275

\bibitem[Semkov et al.(2012)]{semkov}~Semkov, E. H.; Peneva, S. P.; Munari, U.; Tsvetkov, M. K.; Jurdana-\u{S}epi\'{c}, R.; de Miguel, E.; Schwartz, R. D.; Dimitrov, D. P.; Kjurkchieva, D. P. \& Radeva, V. S. 2012, A\&A, 542A, 43

\bibitem[turner(1999)]{turner}~Turner, D. 1999, J. Roy. Astr. Soc. Canada, 93, 187

\bibitem[Zhu et al.(2012)]{zhu}~Zhu, L.-Y.; Zejda, M.; Mikulášek, Z.; Liška, J.; Qian, S.-B. \& de Villiers, S. N. 2012, AJ, 144, 37


\end{thebibliography}
\end{document}